\documentclass[ a4paper, 10pt, twocolumn, superscriptaddress, aps, prd]{revtex4-1} 

\usepackage{amsmath}
\usepackage{amsfonts}
\usepackage{amssymb}
\usepackage{graphicx}
\usepackage{epsfig}
\usepackage{hyperref}
\usepackage{hypernat}
\usepackage{fancyhdr}
\usepackage{natbib}
\usepackage{bm}
\usepackage{float}
\usepackage{lipsum}
\usepackage{color}

\newcommand{\be}{\begin{equation}}
\newcommand{\ee}{\end{equation}}
\newcommand{\ber}{\begin{eqnarray}}
\newcommand{\eer}{\end{eqnarray}}
\newcommand{\bfig}{\begin{figure}}
\newcommand{\efig}{\end{figure}}
\newcommand{\nn}{\nonumber}

\graphicspath{{IDE-PNG_Figs/}{../IDE-PNG_Figs/}}

\begin{document}

\title[Degeneracy between primordial non-Gaussianity and interaction in the dark sector]{Degeneracy between primordial non-Gaussianity and interaction in the dark sector} 

\author{Mahmoud Hashim}
\affiliation{Physics Department, University of the Western Cape, Cape Town 7535, South Africa}

\author{Daniele Bertacca} 
\affiliation{Physics Department, University of the Western Cape, Cape Town 7535, South Africa}

\author{Roy Maartens}
\affiliation{Physics Department, University of the Western Cape, Cape Town 7535, South Africa}
\affiliation{Institute of Cosmology \& Gravitation, University of Portsmouth, Portsmouth PO1 3FX, UK}

\date{\today}

\begin{abstract}
	 If dark energy and dark matter interact via exchange of energy and momentum, then this may affect the galaxy power spectrum on large scales. When this happens, it may be degenerate with the signal from primordial non-Gaussianity via scale-dependent bias. We consider a class of interacting dark energy models and show that the matter overdensity is scale-dependent on large scales. We estimate the effective non-Gaussianity arising  from the large-scale effects of interaction in the dark sector. The signal of dark sector interaction can be disentangled from a primordial non-Gaussian signal by measuring the power at two redshifts.  
\end{abstract}

\maketitle

\section{Introduction} \label{Sec-I}

Deviations from Gaussian initial conditions  offer an important window into the very early universe and a powerful constraint for the mechanism which generated the primordial perturbations. Large primordial Non-Gaussianity (PNG) could be produced within multi-field inflation models, while standard single-field slow-roll models lead to a small level of PNG \cite{Bartolo:2004if}. 
	
PNG can be constrained by observations of the cosmic microwave background \cite{Ade:2013ydc} and the large-scale structure \cite{Giannantonio:2013uqa}. 
In the presence of PNG, the formation of dark matter halos on small scales is modulated by the large-scale overdensity,  leading to a scale-dependent  bias on very large scales. 

Forecasts for PNG constraints typically assume the standard concordance model, i.e. with dark energy as the cosmological constant. 
Dynamical dark energy models, i.e. with equation of state $w_x\neq-1$, do not typically introduce significant changes to the power spectrum on large scales (see \cite{Duniya:2013eta} and references therein). However, if the dark energy interacts with dark matter, then there can be significant effects on the matter overdensity on large scales in some models \cite{Valiviita:2008iv, Potter:2011nv, Duniya:2014}. This means that we could misinterpret a large-scale signal as evidence of PNG when in fact it might be a signature of interacting dark energy (IDE). Here we investigate the large-scale effects on the power spectrum from a class of IDE models, and consider how to disentangle this signal from that of PNG. 
	
The paper is organized as follows. In Sec. \ref{Sec-II} we discuss the perturbations in general IDE models and show the scale-dependence of the matter overdensity for a class of simple models. Section \ref{Sec-IV} considers the effects on the power spectrum of IDE and of PNG, identifies the degeneracy between these signals  and illustrates how to break the degeneracy. Conclusions are given in Sec.\ref{Sec-VIII}.

Our fiducial (non-interacting) model is a $w$CDM (i.e. $w_x=\,$const) model, with $\Omega_{m0}=0.32$, $\Omega_{\Lambda 0}=0.68$ and $n_s=0.96$. The IDE models have the same parameters.

\section{IDE dynamics and perturbations} \label{Sec-II}

The transfer of energy density between dark energy and dark matter is not ruled out by current observations (for recent work, see e.g. \cite{Valiviita:2009nu,Salvatelli:2013wra,Xu:2013jma,Costa:2013sva,Yang:2014gza,yang:2014vza,Wang:2014xca, Salvatelli:2014zta}). Baryons, as standard model particles, are excluded from non-gravitational interaction with the dark sector. For simplicity, and since we are not producing observational constraints, we neglect baryonic matter in our analysis, but it is straightforward to include it via the transfer function.

\subsection*{Background dynamics} 

For interacting dark fluids with energy density $\rho_m$ (dark matter) and $\rho_x$ (dark energy), the background continuity equations are (where $A=m,x$)
\be \label{ce}
\rho_A'+3(1+w_A)\rho_A={aQ_A\over{\cal H}}, ~~Q_x=-Q_m.
\ee
Here a prime denotes $d/d \ln a$. The equation of state parameters are $w_A=p_A/\rho_A$ and we assume a constant equation of state for dark energy:
\be
w_m=0,~~w_x=\,\mbox{const}\neq-1.
\ee
In the non-interacting case, this dark energy model is known as $w$CDM.
The rate of energy density transfer to fluid $A$ is  $Q_A$, and the conservation of total energy enforces $Q_x+Q_m=0$. We can rewrite \eqref{ce} in terms of an effective equation of state:
\begin{equation}\label{Eq-3}
\rho_A'+3(1+w^{\text{eff}}_A)\rho_A=0,~~	w^{\text{eff}}_A =  w_A - \frac{a Q_A}{ 3{\cal H} \rho_A}. 
\end{equation}

The Friedmann constraint and evolution equations do not contain interaction terms since they govern the total density and pressure:
\ber
{\cal H}^2&=&{8\pi Ga^2\over 3}(\rho_m+\rho_x),\\
\label{Eq-4}
	{\cal H}' &=&- \frac{1}{2}  (1 + 3w_t)  {\cal H} ,~w_t = \sum\nolimits_{A}\! w_A \Omega_A=w_x\Omega_x,
\eer 
where $w_t$ is the total equation of state.

\subsection*{ Perturbations}  
Scalar perturbations of the flat background metric in Newtonian gauge are given by
\begin{equation}\label{Eq-6}
 	ds^2 = a^2 \left[ -(1 + 2\Phi) d\tau^2 + (1 - 2\Phi) d\mathbf{x}^2 \right],  
\end{equation} 
where $\Phi$ is the gravitational potential (and equal to the curvature perturbation).
The $A$-fluid energy-momentum tensor is
\begin{equation}\label{Eq-7}
T^{\mu}_{A\nu} = (\rho_A + P_A) u^{\mu}_A u_{\nu}^A + P_A \delta^{\mu}_{\nu},
\end{equation} 
where we assume each fluid is a perfect fluid.
Here $u^{\mu}_A$ is the $A$-fluid four-velocity,
\begin{equation}\label{Eq-8}
u^{\mu}_A = a^{-1} \left(1 - \Phi, \partial^i v_A \right),
\end{equation}
where $v_A$ is the peculiar velocity potential.

The covariant form of energy-momentum transfer is given by (we follow the approach of \cite{Valiviita:2008iv})
\begin{equation}\label{Eq-9}
\nabla_{\nu} T^{\mu\nu}_A = Q^{\mu}_A, ~~~Q^{\mu}_A = Q_A u^{\mu} + F^{\mu}_A.
\end{equation}
We have split the energy-momentum transfer 4-vector relative to the total four-velocity, where
\ber 
\label{Eq-10}
&&u^{\mu} = a^{-1} \left(1 - \Phi, \partial^i v_t \right),\\&& (1 + w_t) v_t = \sum\nolimits_{A}\! (1 + w_A)\Omega_A v_A,\\
&& Q_A = \bar{Q}_A + \delta Q_A,~~ u_{\mu} F^{\mu}_A = 0.
\eer 
Here $v_t$ is the total velocity potential, the energy density transfer rate is $Q_A$ and $F^{\mu}_A$ is the momentum density transfer rate, relative to $u^{\mu}$. For convenience, we drop the overbar on the background $\bar{Q}_A$ from now on. 

Then it follows that  
\begin{eqnarray}\label{Eqn-1}
F^{\mu}_A& =& a^{-1} (0, \partial^i f_A),\\
Q^A_0 &=& -a \left[ Q_A(1 + \Phi) + \delta Q_A\right], \\
Q^A_i &=& a\partial_i\left[ f_A + Q_A v_t\right], 
\end{eqnarray}
where $f_A$ is the momentum transfer potential.
Total energy-momentum conservation implies
\begin{equation}\label{Eq-11}
0 = \sum\nolimits_{A}\ Q_A = \sum\nolimits_{A} \delta Q_A = \sum\nolimits_{A}\ f_A.
\end{equation}

The perturbed Einstein equations do not explicitly contain interaction terms, since they govern the total density and velocity perturbations.
The gravitational potential evolves as
\be \label{Eqn-5}
	\Phi' + \Phi  = - \frac{3}{2} {\cal H} \sum\nolimits_A \Omega_A(1+w_A)v_A,
\ee
and the relativistic Poisson equation is
\be \label{rp}
\nabla^2	\Phi =  \frac{3}{2} {{\cal H}} \sum\nolimits_A \Omega_A\Big[ \delta_A-3 {\cal H}(1+w_A)v_A\Big],
\ee
where $\delta_A={\delta\rho_A / \rho_A}$ is the overdensity in Newtonian gauge.
Although there are no explicit interaction terms in \eqref{Eqn-5} and \eqref{rp}, the gravitational potential $\Phi$ and the matter overdensity $\delta_m$ are affected by interaction -- via the perturbed conservation equations [\eqref{Eqn-2}, \eqref{cont} below], which do explicitly contain the interaction.

In the Newtonian gauge, the galaxy bias $b$ is defined by $\delta_g(k,a)=b(a)\delta_m(k,a)$. However, this definition fails on very large scales due to gauge-dependence, and we need to identify the correct physical frame in which the bias is scale-independent on all scales \cite{Bruni:2011ta}. This is the comoving frame, so that a gauge-independent definition of the bias that applies on all (linear) scales is given by $\Delta_g=b\Delta_m$, where $\Delta_m=\delta_m+(\rho_m'/\rho_m)v_m$. 

For this reason, it is convenient to use the comoving overdensities
\begin{equation} \label{Eq-12}
	\Delta_A = \delta_A + \frac{\rho'_A}{\rho_A} v_A.
\end{equation}
In terms of the comoving overdensities,
the Poisson equation becomes 
\ber \label{Eq-13}
\nabla^2	\Phi&=&  \frac{3}{2} {{\cal H}^2} \Big(\sum\nolimits_A \Omega_A \Delta_A - {\cal Q}^{\Phi}\Big),\\
\label{Eq-14}
	{\cal Q}^{\Phi}& =& {a\over \rho_t} \sum\nolimits_{A} Q_A  v_A={a\over \rho_t} Q_x(  v_x-v_m).
\eer
The interaction is now explicitly present through the velocity terms introduced via the comoving overdensities.

The perturbed conservation equations in \cite{Valiviita:2008iv} are given in terms of $\delta_A$. We re-express these in terms of $\Delta_A$ to obtain 
\begin{eqnarray} \label{Eqn-2}  
	&& v'_A + v_A + \frac{c^2_{sA}}{(1 + w_A) {\cal H}} \Delta_A + \frac{\Phi}{{\cal H}} = {\cal Q}_A^v, \\\label{cont}
	&& \Delta'_A - 3 w_A \Delta_A - \frac{k^2}{{\cal H}} (1 + w_A) v_A \nn \\
	&& - \frac{9}{2} {\cal H} (1 + w_A) (1 + w_t) ( v_A - v_t) = {\cal Q}_A^\Delta ,
\end{eqnarray}
where $c_{sA}$ is the sound-speed, i.e. the speed of propagation of fluctuations. For dark matter, $c_{sm}=0$. For dark energy, we choose the sound-speed of a quintessence scalar field \cite{Valiviita:2008iv}, so that $c_{sx}=1$. The source terms on the right encode the effect of interactions, and are given by
\begin{eqnarray}
	{\cal Q}_{A}^v &=& \frac{a}{(1 + w_A) \rho_A {\cal H}} \Big[Q_A \left(v_t -  v_A \right) + f_A \Big],  \\
	{\cal Q}_A^\Delta &=& \frac{a Q_A }{\rho_A}\left[ \frac{Q'_A}{Q_A} - \frac{\rho'_A}{\rho_A} \right] v_A \nn \\
								&& - \frac{a Q_A }{\rho_A} \left[ 3 + \frac{a Q_A}{(1 + w_A) \rho_A {\cal H} }\right] \left(v_t -  v_A \right)\nn \\
								&& - \frac{a}{\rho_A} \left[ 3 + \frac{a Q_A }{(1 + w_A) \rho_A {\cal H}}\right] f_A\nn \\								
								&& + \frac{a Q_A}{\rho_A} \left[3 (1 + w_A) + \frac{a Q_A}{\rho_A {\cal H}}\right] v_A + \frac{a }{\rho_A {\cal H}}\,\delta Q_A \nn\\
								&& - \frac{a Q_A}{\rho_A {\cal H}} \left[ \frac{c^2_{sA}}{(1 + w_A)} + 1\right] \Delta_A + 2 \frac{a Q_A}{\rho_A {\cal H}}\, \Phi.
\end{eqnarray}

\subsection*{A simple model of IDE}

\begin{figure*}
\includegraphics[scale = 0.45]{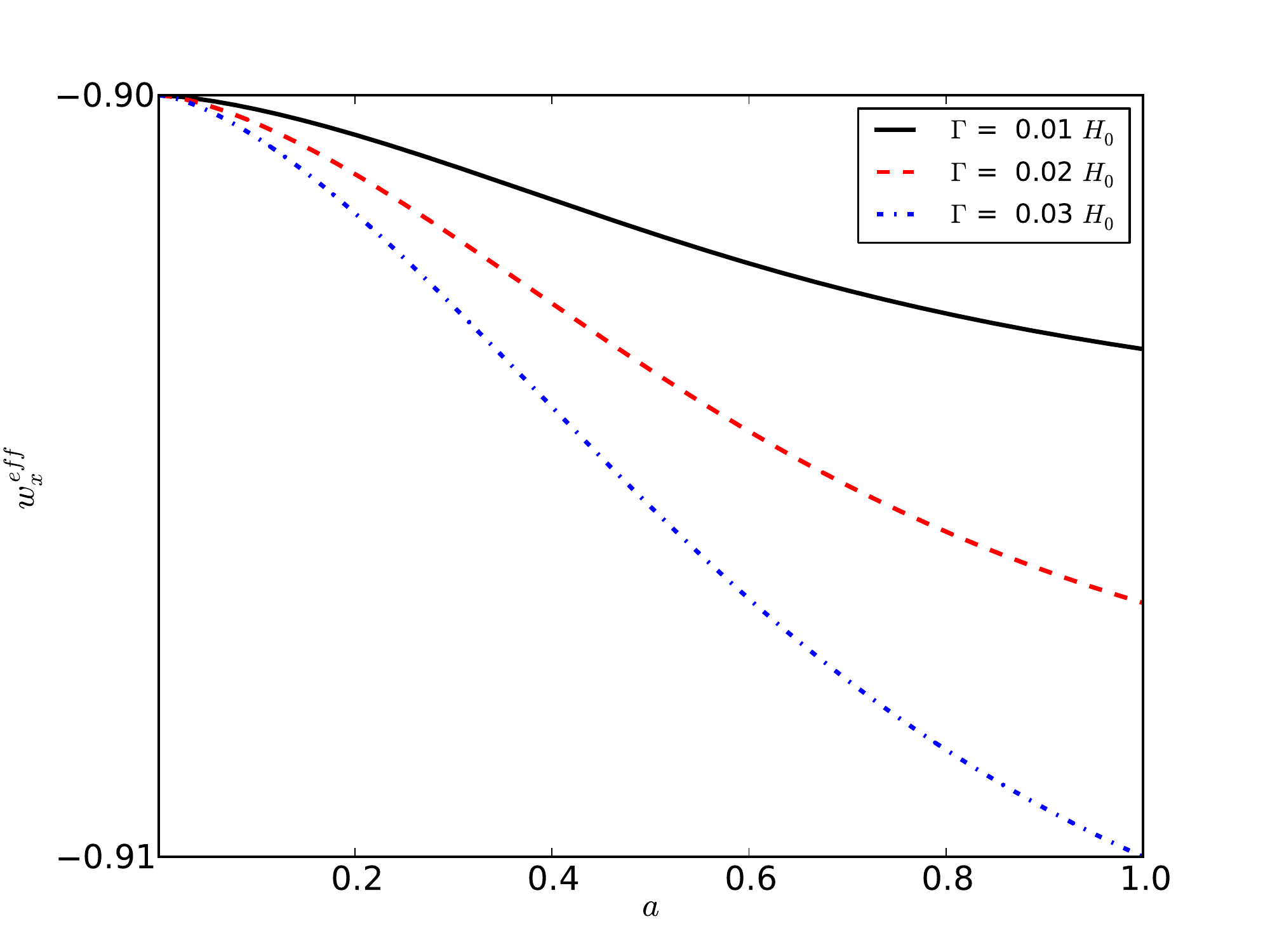}~
\includegraphics[scale = 0.45]{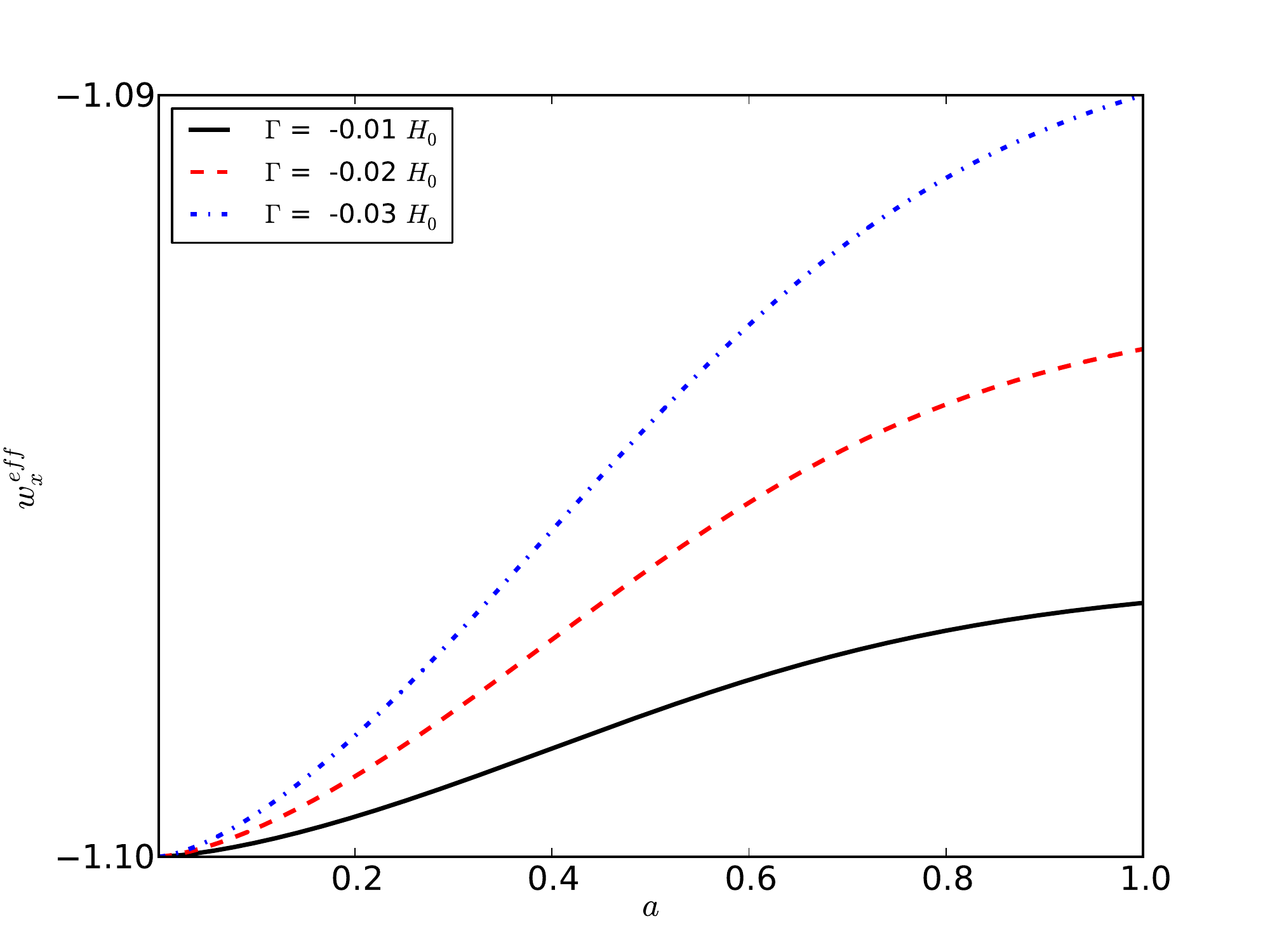}
\caption{{\em Left:} Evolution of effective  dark energy equation of state, $w^{\text{eff}}_x$, with $w_x= -0.9$ and for different $\Gamma>0$, in the IDE model  \eqref{Eq-15}. The $\Gamma=0$ limit is $w^{\text{eff}}_x=w_x=-0.9$. {\em Right:} The $\Gamma<0$ case, with $w_x=-1.1$.} \label{Fig-I}
\end{figure*}

Dark sector interactions are not ruled out observationally and are theoretically plausible, given the unknown nature of the physics of the dark sector. We do not have guidance from fundamental physics for either the nature of dark energy, or the form of a possible interaction between it and dark matter. Here we choose a simple model of dark energy and a simple interaction model, each with only a single parameter.

We adopt the interaction model of \cite{Clemson:2011an}, defined covariantly by
\begin{equation} \label{Eq-15}
Q^{\mu}_x =  - Q^{\mu}_m = \Gamma \rho_x u^{\mu}_x, 
\end{equation}
where $\Gamma$ is a constant interaction rate. Since the dark energy has $w_x=\,$const, we call this model $\Gamma w$CDM. 

In the background, \eqref{Eq-15} gives
\begin{equation}\label{Eq-16}
Q_x = \Gamma \rho_x=-Q_m,
\end{equation}
and for the perturbations,  
\begin{equation}\label{Eq-17}
\delta Q_x = -\delta Q_m = \Gamma \rho_x \delta_x, ~~ f_x = - f_m =\Gamma \rho_x (v_x - v_t). 
\end{equation}

Since $Q^\mu_A$ is parallel to $u^\mu_x$, there is no momentum transfer in the dark energy frame.
This means there is momentum transfer in the dark matter frame, so that the dark matter velocity $v_m^i$ does not obey the same Euler equation as the galaxies, and there is consequently a velocity bias \cite{Koyama:2009gd}. The alternative model considered in \cite{Clemson:2011an} has $Q^\mu_A$ parallel to $u^\mu_m$, without momentum transfer in the dark matter frame and thus with no velocity bias. We have checked that our results are not qualitatively different for this alternative model.

There are two cases for the $\Gamma w$CDM model:
\begin{itemize}
\item
$\Gamma>0$ -- which represents  a {\em transfer of energy density from dark matter to dark energy}, with transfer  rate $\Gamma$.

Stability of this model requires \cite{Clemson:2011an} $w_x>-1$. 

\item
$\Gamma<0$ -- which represents the {\em decay of dark energy to dark matter}, with decay rate $|\Gamma|$.

Stability of this model requires \cite{Clemson:2011an} $w_x<-1$. 
\end{itemize}

From \eqref{Eq-3} and \eqref{Eq-16}, we see that $w^{\rm eff}_x=w_x-a\Gamma/(3{\cal H})$. Then it follows that $w^{\rm eff}_x<w_x$ when $\Gamma>0$ and $w^{\rm eff}_x>w_x$ when $\Gamma<0$. This behaviour is illustrated in
Fig. \ref{Fig-I}. The effects of interaction grow with time, as dark energy becomes significant and then dominant. It is clear that $|\Gamma|/H_0<1$ is required to avoid a background evolution that will be ruled out by distance measurements.  Here are we are not concerned with precise limits on $\Gamma$ (see \cite{Clemson:2011an} for these).

\subsection*{Initial conditions}

At decoupling, we assume that the dark fluids are adiabatic and have equal peculiar velocities:
\begin{equation} \label{Eq-18}
S_{mx}\big|_d \equiv \left(\frac{\delta\rho_m}{\rho'_m}   - \frac{\delta\rho_x}{\rho'_x}  \right)_d= 0, ~~~  v_{md} = v_{xd}.
\end{equation}
By \eqref{Eq-12}, this implies
\begin{equation}\label{Eq-43}
\left(\frac{\rho_m}{\rho'_m}\right)_d \Delta_{md} = \left(\frac{\rho_x}{\rho'_x}\right)_d \Delta_{xd}.
\end{equation}
Using the Poisson equation \eqref{Eq-13}, we get 
\begin{eqnarray} \label{Eqn-7}
    \Delta_{md} &=& - \frac{2}{3}  \left(\frac{k}{{\cal H}_d}\right)^2   \frac{(1 + \mu)}{\Omega_{md}} \Phi_d, \\
    \Delta_{xd} &=& - \frac{2}{3}  \left(\frac{k}{{\cal H}_d}\right)^2 \frac{\mu}{\Omega_{xd} } \Phi_d,
\end{eqnarray}
where we have defined
\begin{eqnarray} \label{Eqn-8}
\mu &=& \left(\frac{\rho'_x}{\rho'_m} \right)_d \left[1 - \left(\frac{\rho'_x}{\rho'_m}\right)_d \right]^{-1} \ll 1.
\end{eqnarray}
We find that $\mu \sim 10^{-9}$ for   $|\Gamma|/H_0\lesssim 0.03$ and $w_x=-0.9$. For $w_x = -1.1$, $\mu \sim -10^{-11}$.

 \begin{figure*}
\includegraphics[scale = 0.44]{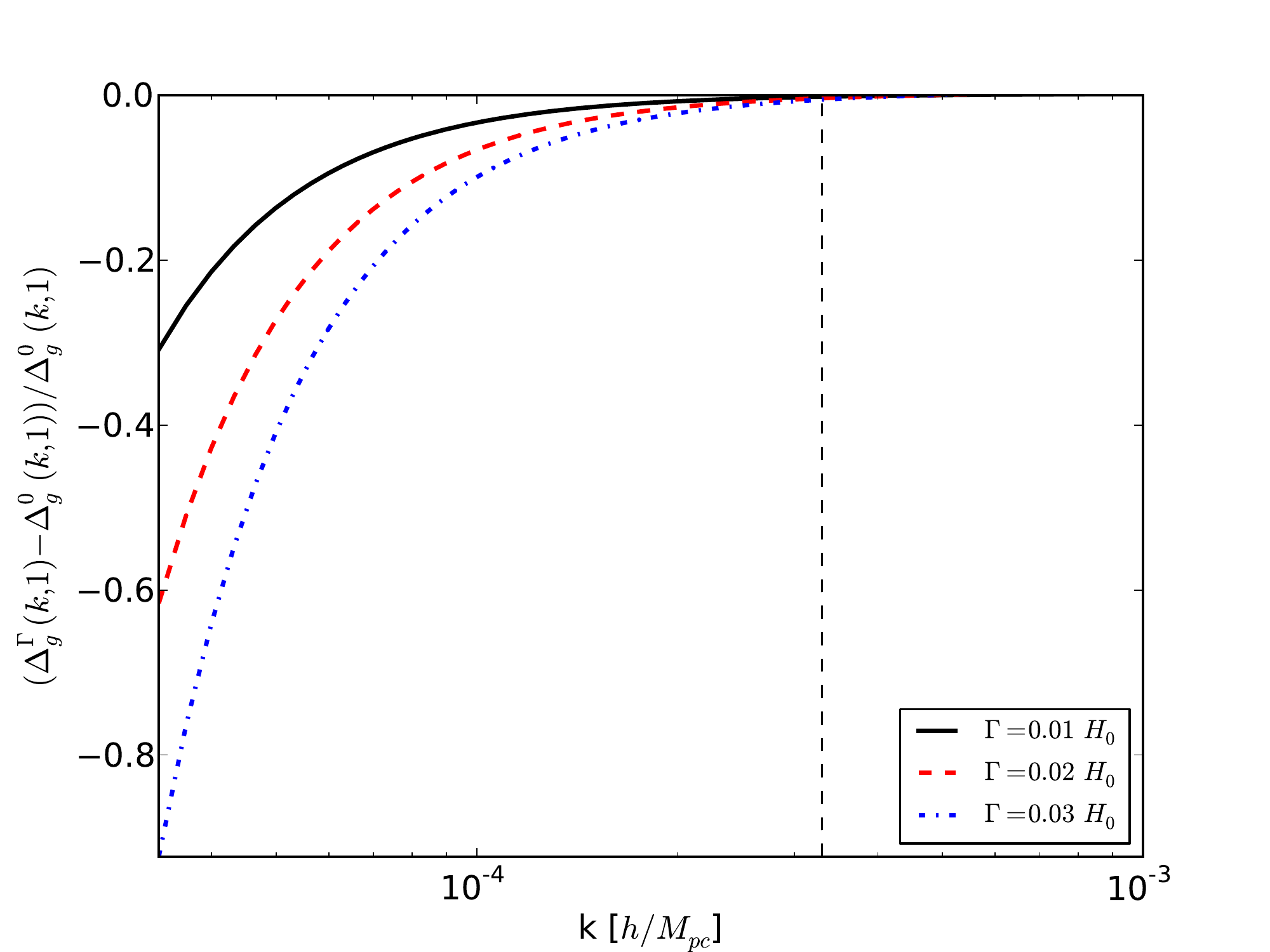}~
\includegraphics[scale = 0.44]{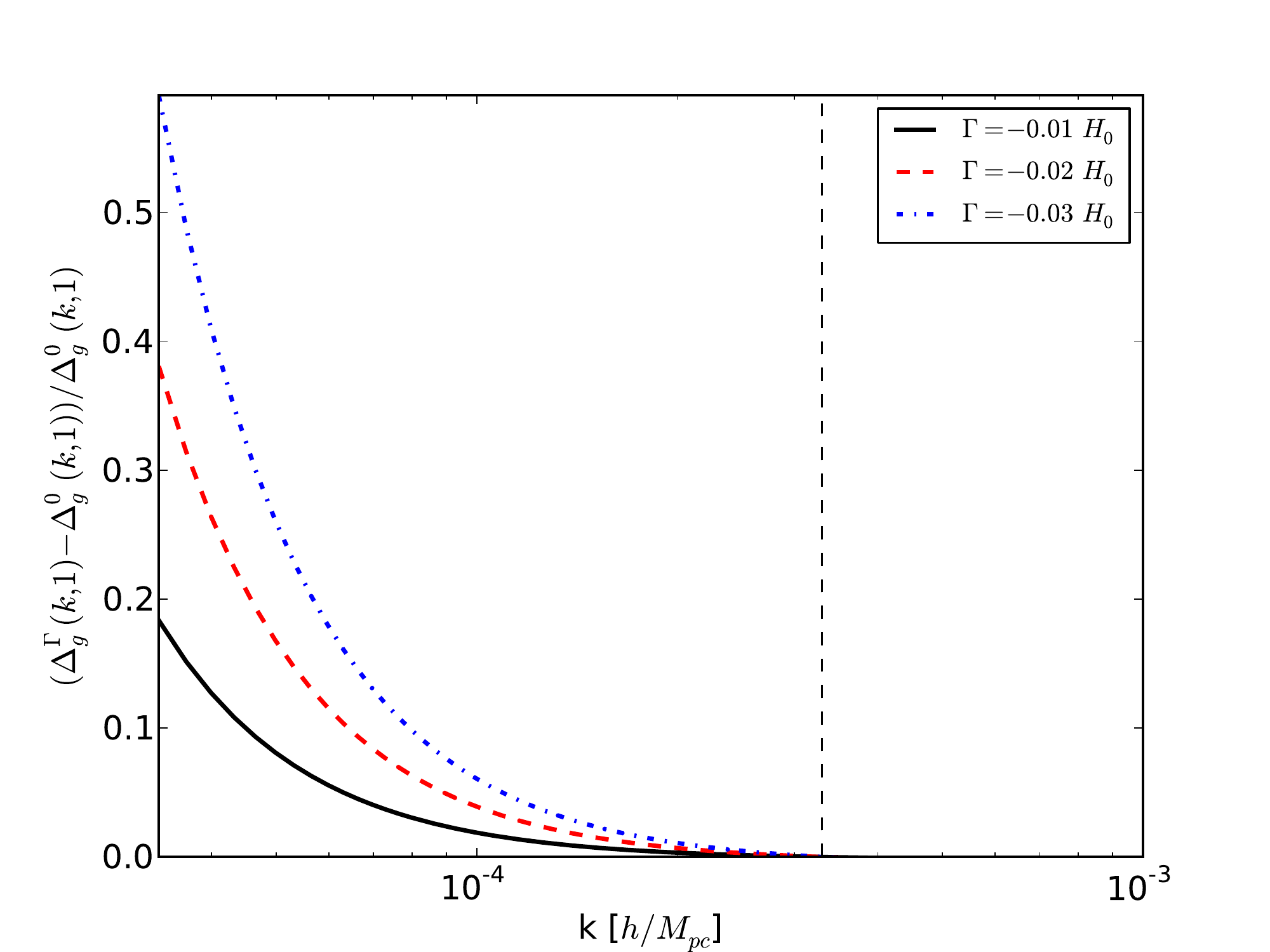}
\caption{{\em Left:} Relative galaxy overdensity [see \eqref{rgr}] at $a=1$ with dark sector interactions, for different  $\Gamma>0$ and with $w_x=-0.9$. The vertical dashed line is the Hubble scale, $k=H_0$. The $\Gamma=0$ limit is the horizontal line through 0.  We used $b(1)=2$. {\em Right:} The $\Gamma<0$ case, with $w_x=-1.1$. 
} \label{Fig-III}
\end{figure*}
\begin{figure*}
\includegraphics[scale = 0.45]{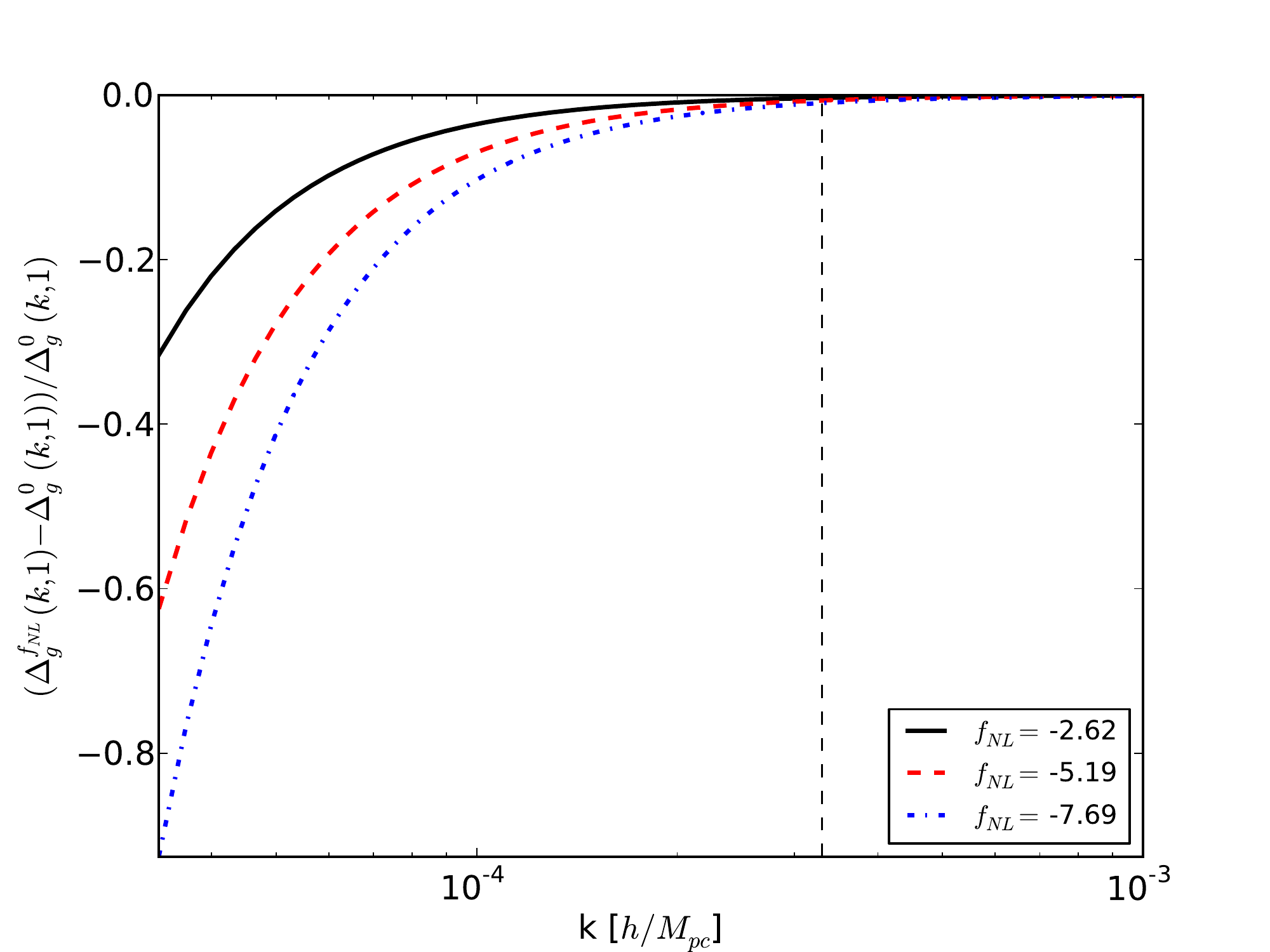}~
\includegraphics[scale = 0.45]{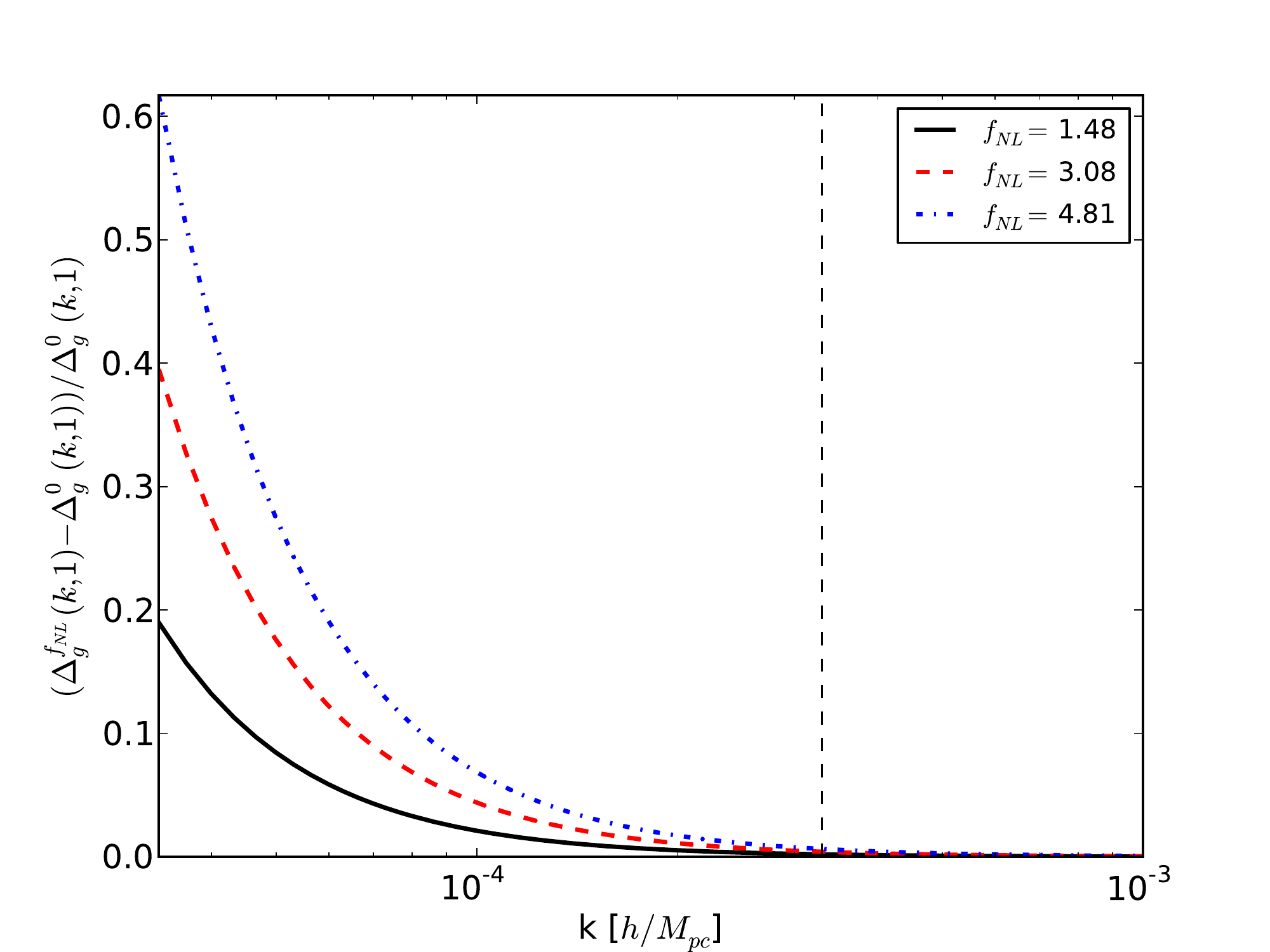}
\caption{The relative galaxy overdensity as in Fig. \ref{Fig-III} but for the case of PNG, with $f_{\rm NL}<0$ ({\em left}) and $f_{\rm NL}>0$ ({\em right}). } \label{Fig-IV}
\end{figure*}

The gravitational potential at decoupling is related to the primordial potential as follows:
\begin{eqnarray} \label{Eq-44}
\Phi_d(k) &=& \frac{9}{10} T(k) \Phi_p(k), \\
\label{Eq-45}
\Phi_p(k) &=& A 
\left( \frac{k}{H_0}\right)^{(n_s-4)/2}, 
\end{eqnarray}
where $T$ is the transfer function ($\to 1$ on very large scales), $n_s$ is the spectral index of the primordial spectrum and  $A$ is an amplitude determined by the primordial curvature perturbation.  

\begin{figure*}[t]
\includegraphics[scale = 0.45]{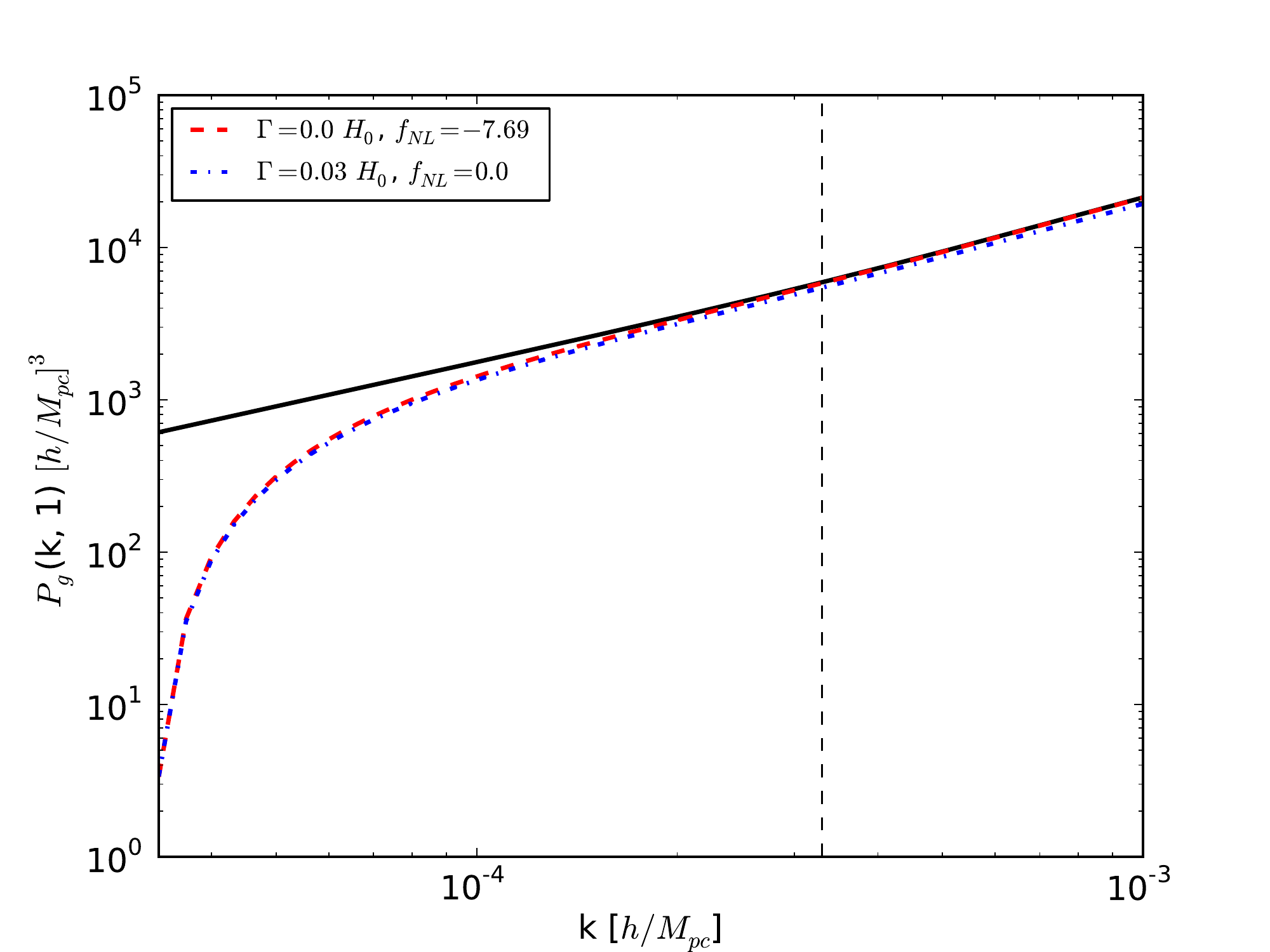}~
\includegraphics[scale = 0.45]{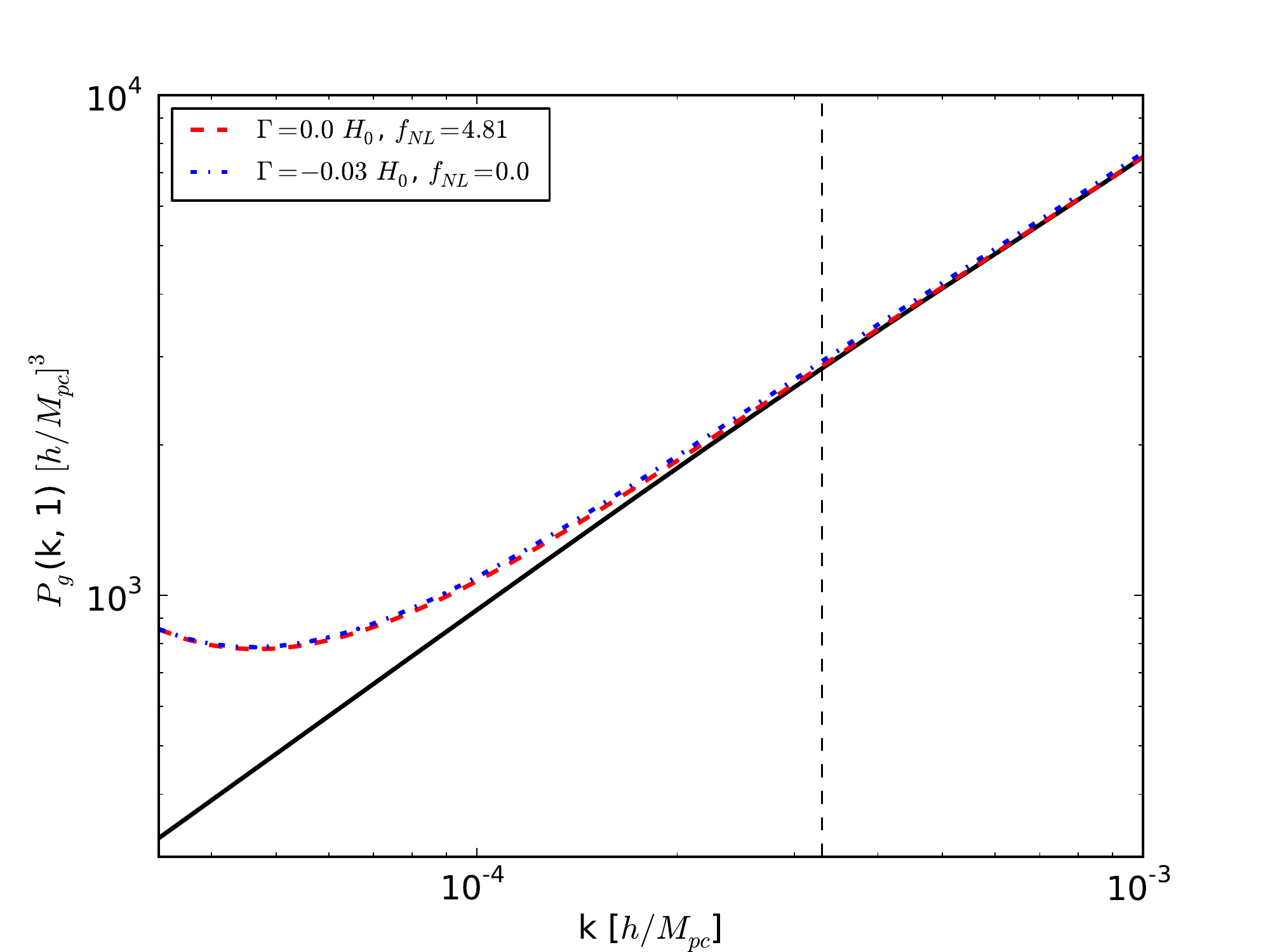}
\caption{{\em Left:} Galaxy power spectrum $P_g$ at $a=1$ for an IDE model with $\Gamma=0.03H_0$ and for a PNG model with $f_{\rm NL}=-7.69$. The black (solid) line is the fiducial $w$CDM model without interaction or PNG. We set $b(1)=2$ and $w_x=-0.9$. {\em Right:} For
$\Gamma=-0.03H_0$ and  $f_{\rm NL}=4.81$, with $w_x=-1.1$.} \label{Fig-V}
\end{figure*}

We can neglect dark energy and the interaction at decoupling provided that $\mu\ll1$ and $|\Gamma|/H_0 \lesssim 0.03$. This is equivalent to assuming that the universe at decoupling is well described as matter-dominated, and it implies that $\Phi'_d = 0$. Then from  \eqref{Eqn-5}, \eqref{Eq-14}  and \eqref{Eq-18}, we find
\ber \label{Eq-46}
 	v_{md} &=&   - \frac{2}{3(1 + w_x \Omega_{xd}){\cal H}_d } \Phi_d=v_{xd},\\ 
{\cal Q}^\Phi_d &=& 0.
\eer

\subsection*{Growth functions} \label{Sec-III}

The potential growth function $D_\Phi$ is defined by
\begin{equation}\label{Eq-23}
\Phi(k,a) = \frac{D_{\Phi}(k,a)}{a} \Phi_d(k),
\end{equation}
so that $D_{\Phi d} = a_d$.

We  define dark matter and dark energy growth functions  
\begin{eqnarray} \label{Eqn-3} 
 D_m(k,a) &= & \frac{\Delta_m(k,a)}{\Delta_{md}(k)}  a_d, \\
 D_x(k,a) &=& \frac{\Delta_x(k,a)}{\Delta_{xd}(k)}  a_d,
\end{eqnarray}
where we normalize at decoupling. Then it follows from the Poisson equation \eqref{Eq-13}  that
\ber
D_m &=& {\Omega_{md} \over \Omega_{m}(1+\mu)} \Bigg[{a_d {\cal H}_d^2\over a{\cal H}^2} D_\Phi \nonumber\\ \label{dm}
&&~-\mu{\Omega_x \over \Omega_{xd}}D_x -B {a_d {\cal H}_d^2\over T(k)k^{n_s/2}} {\cal Q}^\Phi  \Bigg],\\
{\cal Q}^\Phi &=& a\Gamma\Omega_x\left(v_x-v_m\right). \label{qp}
\eer
Here $B=5H_0^{(4-n_s)/2}/(3A)$ is a constant. 
In the limiting case of the concordance model $\Lambda$CDM ($\Gamma=0, w_x=-1$), we have $\mu=0=D_x$ and ${\cal H}^2\Omega_m = a^{-1} H_0^2\Omega_{m0}$. Thus \eqref{dm} 
recovers the $\Lambda$CDM relation $D_m=D_\Phi$. In $\Lambda$CDM, the matter growth function is scale-independent, $D_m=D_m(a)$. This also holds approximately for the non-interacting $w_x=\,$const models, $w$CDM. 
The effect of dark sector interactions on the growth of the comoving matter overdensity  is encoded in the ${\cal Q}^\Phi$ term \eqref{qp}.

\section{Large-scale power in IDE} \label{Sec-IV}

We show below that dark sector interactions in our model do lead to a growth or decrease of matter power on large scales -- which is similar to the effect of PNG on the galaxy power spectrum. This  illustrates the point that if we are unaware of the possibility of IDE, then a detection of PNG from the galaxy power spectrum could in fact be a signal of dark sector interaction with Gaussian primordial perturbations. We need to be able to distinguish the two possibilities, i.e. to break the potential degeneracy between the signals of IDE and PNG in the large-scale galaxy power. First we need to characterize the galaxy power in the two scenarios.

\subsection*{Galaxy overdensity in IDE and PNG}

In the absence of PNG, the galaxy overdensity in $\Gamma w$CDM is related to the matter overdensity  on linear scales by 
\be\label{del1}
\Delta^\Gamma_g(k,a) = b(a) \Delta_m(k,a),
\ee
where $b$ is the scale-independent bias. We have introduced a $\Gamma$ superscript to distinguish this galaxy overdensity from the non-interacting PNG case.

\begin{figure*}
\includegraphics[scale = 0.45]{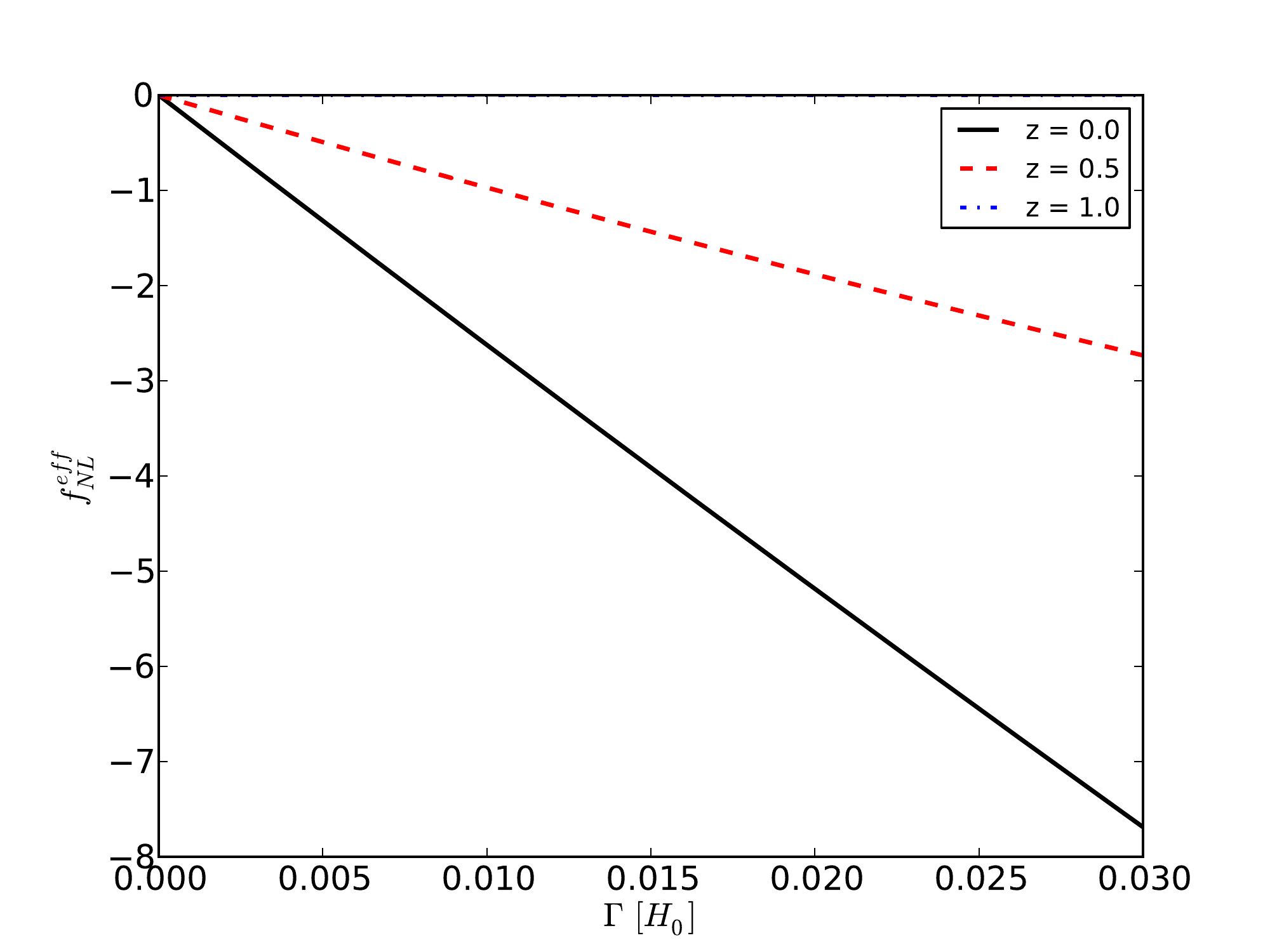}~ 
\includegraphics[scale = 0.45]{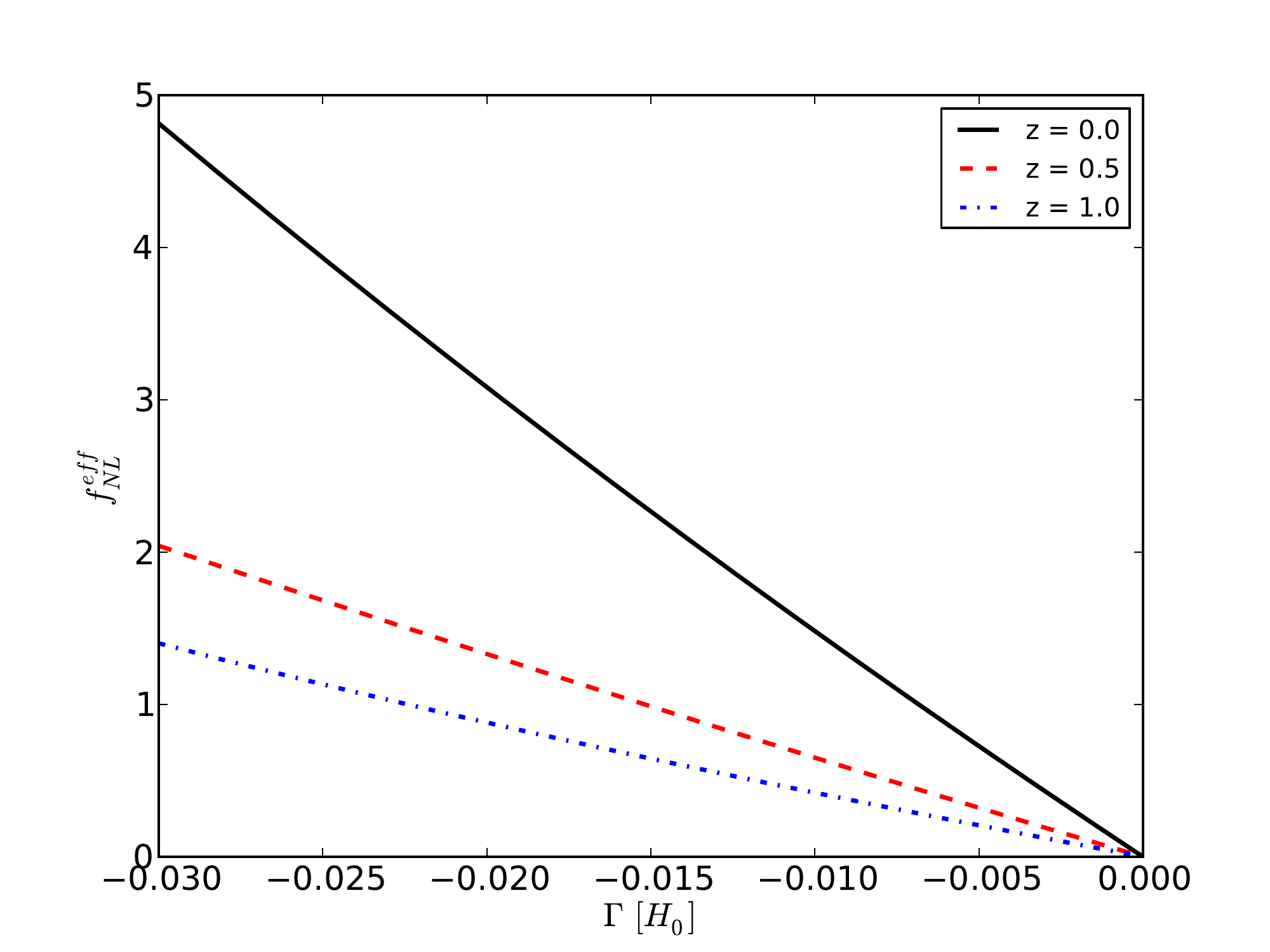} 
\caption{{\em Left:} Effective PNG parameter $f^{\text{eff}}_{\rm NL}$ corresponding to interaction rate $\Gamma>0$ at different redshift values. We take $w_x=-0.9$. {\em Right:} For
$\Gamma<0$ and $w_x=-1.1$.} \label{Fig-VII}
\end{figure*} 

In the presence of PNG, the bias becomes scale-dependent on large scales \cite{Dalal:2007cu}:
\ber\label{Eq-31-1}
b(a) &\to& b(a) + \Delta b(k,a),\\
\label{Eq-32}
\Delta b &=& 3 f_{\rm NL} (b - 1)\frac{ \delta_c H_0^2 \Omega_{m0}}{k^2 T D_m},
\eer
where $f_{\rm NL}$ is the local PNG parameter and $\delta_c$ is the critical overdensity for halo collapse. On very large scales, where $T\to1$, we have $\Delta b \propto f_{\rm NL}k^{-2}$. Equation \eqref{Eq-32} is derived in $\Lambda$CDM but can also be applied to $w$CDM with the replacement
\be
D_m(a) \to (1+\mu)D_m(k,a),
\ee
where $\mu$ is given by \eqref{Eqn-8} and $D_m(k,a)$ is the growth function for non-interacting dark energy (i.e., $\Gamma=0$). This replacement makes only a small change provided $w_x$ is close to $-1$ and $c_{sx}=1$. Thus for non-interacting $w$CDM with PNG
\be\label{del2}
\Delta^{f_{\rm NL}}_g (k,a) = b(a)\Big[1+ {\Delta b(k,a) \over b(a)}\Big] \Delta_m^0(k,a),
\ee
where $\Delta_m^0$ is the matter overdensity in $w$CDM.
We will call this model $f_{\rm NL}w$CDM.

\subsection*{Comparing the galaxy power }

Now we investigate whether the large-scale behaviour is qualitatively similar in the two cases, i.e. the 
$\Gamma w$CDM and $f_{\rm NL}w$CDM models. In order to do this, we define for each model the galaxy overdensity relative to $w$CDM, i.e.,
\be\label{rgr}
{\Delta_g^\Gamma(k,a)-\Delta_g^0(k,a) \over \Delta_g^0(k,a)} ~\mbox{and}~ {\Delta_g^{f_{\rm NL}}(k,a)-\Delta_g^0(k,a) \over \Delta_g^0(k,a)} ,
\ee
where $\Delta_g^0(k,a)=b(a) \Delta_m^0(k,a)$ denotes the $w$CDM galaxy overdensity ($\Gamma=0= f_{\rm NL}$). These relative overdensities are shown in 
Figs. \ref{Fig-III} and \ref{Fig-IV}. 

\subsection*{IDE mimics PNG}

Figures \ref{Fig-III} and \ref{Fig-IV} confirm that the effects of $\Gamma$ and of $f_{\rm NL}$ are qualitatively similar, giving a growth ($\Gamma < 0,f_{\rm NL}>0$) or suppression ($\Gamma>0,f_{\rm NL}<0$) on super-Hubble scales. The effect is stronger as $|\Gamma|$ or $|f_{\rm NL}|$ are increased. By comparing the galaxy power spectra, we find numerically the effective PNG parameters that correspond most closely to $|\Gamma|/H_0=0.03$. The correspondence is confined to scales that are not too far beyond the Hubble radius, since this regime is well outside the reach of observations. 

In Fig. \ref{Fig-V}, we compare the resulting  galaxy power spectra, where 
$P_g(k,a) = \langle |\Delta_g(k,a)|^2\rangle$.

Figure \ref{Fig-V} indicates that we can successfully extract an effective PNG parameter when $|\Gamma|/H_0= 0.03$. We extend this 
 over the range $|\Gamma|/H_0< 0.03$ to produce a curve of  the effective PNG parameter against $\Gamma$. We do this for a range of redshifts,
and the results are shown in Fig. \ref{Fig-VII}. To account for the redshift evolution of the (Gaussian) bias on linear scales, we adopt the ansatz $b=b_0\sqrt{1+z}=b_0a^{-1/2}$ with $b_0=2$.

\subsection*{Breaking the degeneracy between IDE and PNG}

Figure \ref{Fig-VII}, shows a key feature:
\begin{itemize}
\item
As redshift increases,  the value of $|f^{\text{eff}}_{\rm NL}|$  decreases, approaching zero at redshifts $z \gtrsim 1$. This follows since the dark sector interaction only begins to have an effect on galaxy power at late times. By contrast,
the PNG signal is `frozen' into the power spectrum at primordial times so that $f_{\rm NL}$ is independent of redshift.

\end{itemize}

This feature should be generic for IDE models that cause large-scale deviations in the power spectrum.
It is exactly what allows us to break the degeneracy between PNG and IDE using the galaxy power spectrum. If we establish a value of $f^{\text{eff}}_{\rm NL}$ at redshift $z=0$, then we can compare the observed power at another redshift, e.g. $z=0.5$, with that predicted by PNG with $f_{\rm NL}=f^{\text{eff}}_{\rm NL}$. Significant disagreement indicates that the large-scale signal is not due to PNG, but could be a smoking gun for dark sector interaction.

For our IDE model, the relationship between $f^{\text{eff}}_{\rm NL}$ and $\Gamma$ can be estimated analytically as follows. We take the limit $k\ll{\cal H}$, so that \eqref{del2} implies $\Delta_g^{f_{\rm NL}}\to \Delta b\Delta^0_m$. In the Poisson equation \eqref{Eq-13}, we neglect $k^2\Phi$ to obtain  $\Delta_m\to {\cal Q}^\Phi/\Omega_m$. Thus by \eqref{del1}, $\Delta_g^\Gamma\to b{\cal Q}^\Phi/\Omega_m$. Then we can write (for $k\ll{\cal H}$)
\be
\Delta_g^{f^{\text{eff}}_{\rm NL}} \approx \Delta_g^\Gamma ~\Rightarrow~
\Delta b\Delta^0_m \approx b {{\cal Q}^\Phi \over \Omega_m}.
\ee
From \eqref{Eqn-3} we have $\Delta^0_m=D^0_m\Delta^0_{md}/a_d$. Then \eqref{Eqn-7}, \eqref{Eq-44} and \eqref{Eq-32} (with $T=1$) imply
\[
f^{\text{eff}}_{\rm NL}\approx\Bigg[ {5a_d\over9A}{a\Omega_{md}{\cal H}_d^2 \over \Omega_{m0}H_0^2}{b\over(b-1)\delta_c}{\Omega_x\over \Omega_m}\!\left(\!{k\over H_0}\!\right)^{\!\!(4-n_s)/2}\! \!(v_x-v_m) \Bigg]\!\Gamma
\]
\be\ee
Clearly $f^{\text{eff}}_{\rm NL}$ is in general redshift dependent, and this is confirmed by Fig. \ref{Fig-VII}.

\section{Conclusion} \label{Sec-VIII}

We have shown that for a simple class of models, interaction in the dark sector causes a growth or suppression of matter power on very large scales, relative to the non-interacting case. Furthermore, these large-scale deviations can be approximately mimicked by PNG in a non-interacting model. This raises a potential problem for attempts to constrain PNG through the galaxy power spectrum -- such attempts could be confused by interaction in the dark sector. One way to break this degeneracy is by looking at the power spectra at two redshifts. If the two effective parameters $f^{\text{eff}}_{\rm NL}$ are not equal, then this is a strong indication of IDE. 

There are serious obstacles to the observability of any effect that arises only on very large scales -- including IDE effects and PNG. The fundamental problem is cosmic variance, which grows on large scales and typically swamps any small signal. The current state of the art in constraining PNG via galaxy surveys \cite{Giannantonio:2013uqa} is unable to detect $|f_{\rm NL}| \lesssim 20$, and Planck has already placed the constraint $|f_{\rm NL}| \lesssim 10$. In the IDE model that we investigate here, $f^{\text{eff}}_{\rm NL}$ is in the range compatible with Planck, and therefore not currently detectable. In order to overcome the problem of cosmic variance in galaxy surveys, we need either three-dimensional data (i.e., the power spectrum measured over a significant range of redshifts) or the application of the multi-tracer method \cite{Seljak:2008xr,McDonald:2008sh}, or both. (The multi-tracer method requires that we have two or more different tracers of the underlying dark matter overdensity.) Future surveys such as Euclid and the SKA will be needed in order to detect PNG or large-scale IDE effects at the level
  $|f^{\text{eff}}_{\rm NL}|\lesssim 10$ considered here.

There is an important further point about  observability on very large scales: on these scales, there are general relativistic corrections to the standard power spectrum, which are also potentially degenerate with PNG \cite{Bruni:2011ta,Jeong:2011as,Yoo:2012se, Raccanelli:2013dza}. Therefore one needs to include the relativistic effects in any analysis of PNG, as in \cite{Bruni:2011ta,Jeong:2011as,Yoo:2012se, Raccanelli:2013dza}. The same applies to an anlysis of IDE on very large scales. The relativistic effects for our IDE model and others are investigated in \cite{Duniya:2014}.

\[\]
{\bf Acknowledgements:} 
This work is supported by the South African Square Kilometre Array Project and the South African National Research Foundation. RM is also supported by the UK Science \& Technology Facilities Council (grant no. ST/K0090X/1). 

\bibliographystyle{apsrev4-1}  
\bibliography{IDE-PNG-22oct}

\end{document}